\documentclass{ws-mpla}
\usepackage[super]{cite}
\usepackage{graphicx}
\begin{document}

\markboth{G. L.~Klimchitskaya \& V. M.~Mostepanenko}
{Recent Measurements of the Casimir Force}

\catchline{}{}{}{}{}

\title{\uppercase{Recent measurements of the Casimir force:
Comparison between experiment and theory }
}

\author{\uppercase{G.~L.~Klimchitskaya${}^{1,2}$
{\lowercase{and}}
 V.~M.~Mostepanenko${}^{1,2,3}$}}

\address{${}^1$Institute of Physics, Nanotechnology and
Telecommunications, Peter the Great \\Saint Petersburg
Polytechnic University, Saint Petersburg, 195251, Russia\\
${}^2$Central Astronomical Observatory at Pulkovo of the
Russian Academy of Sciences, \\Saint Petersburg,
196140, Russia\\
${}^3$Kazan Federal University, Kazan, 420008, Russia\\
g\_klimchitskaya@mail.ru, vmostepa@gmail.com}

\maketitle

\pub{Received 1 September 2019}{Revised 14 September 2019}

\begin{abstract}
We consider experiments on measuring the Casimir interaction which have
been performed in the last four years. The emphasis is made on
measuring differences in the Casimir pressures under a transition of the plate
metal from normal to superconducting state and on the Casimir metrology
platform using a commercial micromechanical sensor. In both cases several
problems in the comparison between experiment and theory are discussed.

\keywords{Casimir effect; superconductors; micromechanical sensor.}
\end{abstract}

\ccode{PACS Nos.: 12.20.Fv, 12.20.Ds }

\section{Introduction}

Measurements of the Casimir force, which have been actively made in many
laboratories starting in 1997 (see Refs.~\citen{1} and  \citen{1a}
for a review), attract
much attention in connection with their importance for fundamental physics
and prospective technological applications. This is particularly true with
respect to the diversified experiments performed after the Third Casimir
Symposium which had happened in 2015. Two of them were devoted to measurements
of the Casimir force in a gaseous or liquid media\cite{2,3} (see also
Ref.~\citen{4} investigating sensitivity and accuracy of Casimir force
measurements in air). Several experiments needed for future applications of
the obtained results in nanotechnology were devoted to measurements of the
Casimir force in an optomechanical cavity,\cite{5} silicon carbide
systems,\cite{6} and between silicon nanostructures.\cite{7}
An implementation of special techniques, such as Ar-ion and UV cleaning,
 in the laboratory setups allowed
to reduce detrimental electrostatic effects 
which plague the investigation of
Casimir forces.\cite{8,9} Some progress has been also reached in measuring
the Casimir pressure between two parallel plates at separations of a few
micrometers (see the first results\cite{10} and proposed
improvements\cite{11,12}).

The greatest breakthrough, however, was reached in measuring the difference
in Casimir forces between a Ni-coated sphere and either a Ni or Au strips of
the plate covered with an Au overlayer.\cite{13} In this experiment,
the difference between theoretical predictions of two competing theories,
one taking into account (the Drude model approach) and 
other disregarding (the
plasma model approach) the relaxation of free electrons, 
is by a factor of
1000.  As a result, the plasma model approach was found to be consistent
with the measurement data and the Drude model approach was conclusively excluded
be the data\cite{13} (at least at separations below $1~\mu$m).
This finally confirmed the results of previous experiments\cite{1,1a} where
the difference in theoretical predictions of the two approaches was only a
few percent of the measured force.

Here, we discuss problems in the comparison between experiment and theory
in two more recent measurements of the Casimir force. In Sec.~2, an experimental
investigation of the Casimir force under a 
phase transition of plate metal from
normal to superconductor state\cite{14} is considered. Section~3 is devoted
to the Casimir metrology platform using a commercial microelectromechanical
sensor.\cite{15} In Sec.~4, the reader will find our conclusions.

\section{The Casimir Force under a Phase Transition from Normal to
Superconductor State}

The first investigation of the Casimir pressure between two parallel plates
made of superconducting metal (Al) was performed by means of on-chip
optomechanical sensor.\cite{14} In so doing, one of the Al plates was
attached to the movable mirror of an optical cavity. Due to the 
variation in the Casimir pressure, the separation distance between the Al
plates should change resulting in the change of the cavity length and
respective shift of its resonance frequency. Measurements of the expected
frequency shift have been performed with decreasing temperature starting
from 100\,K to 0.01\,K at the separation between the plates
$100\pm 10\,$nm (larger gaps have also been tested).
However, no frequency shift was observed when cooling the plates through
their critical temperature $T_c\approx 1.3\,$K below which Al becomes
superconducting.\cite{14}
This means that up to the measurement errors no change in the Casimir
pressure $\Delta P$ was observed after a transition of the plate metal to
the superconducting state.

The motivation for performing this experiment was that a 
phase transition of the
plate metal into a superconducting state affects its reflectivity properties
at frequencies below $k_BT_c/\hbar$ (where $k_B$ is the Boltzmann constant)
and should also change the magnitude of the Casimir force.\cite{14}
The hope was expressed\cite{14} for the possibility to distinguish between the
theoretical predictions of the Drude and plasma model approaches using
this effect. This hope was based on the theoretical results\cite{16,17}
investigating different approaches to the description of the Casimir force
between metals in a superconducting state (see also recent Ref.~\citen{18}).
For $T>T_c$ both the plasma and Drude model approaches have been used
to calculate the Casimir pressure.
For $T<T_c$ a superconducting metal was described either by the plasma model
(as is suggested in the classical textbook\cite{19}) or by the
phenomenological Mattis-Bardeen dielectric permittivity\cite{20} which is
smoothly joined with the permittivity of the Drude model at $T=T_c$.
If the plasma model is used, the zero frequency shift and, respectively,
$\Delta P=0$ are predicted at  $T<T_c$. When using the Mattis-Bardeen model,
$\Delta P$ was shown again equal to zero when just passing through $T_c$,
but being a decreasing  function with further decrease in $T$.

Thus, the experimental results obtained so far\cite{14} are in favor of the
plasma model. One should note also that the phenomenological Mattis-Bardeen
dielectric permittivity at $T<T_c$ has a $\delta$-function term of the
order $\delta(\omega)/\omega$ at zero frequency.\cite{20}
Permittivities of this kind cannot be analytically continued to the upper
plane of complex frequency, do not satisfy the Kramers-Kronig relation\cite{21}
and, thus, scarcely can be used for making reliable predictions concerning the
behavior of the Casimir force at $T<T_c$.

\section{Casimir Metrology Platform}

In Ref.~\citen{15} the capacitive microelectromechanical inertial sensor
was used for measuring the Casimir force between an Ag-coated microsphere
and an Au-coated silicon plate in ambient conditions at room temperature.
Measurements were performed within the separation region from 50\,nm to
$1\,\mu$m. The electrostatic calibration has been done as described in
previous literature.\cite{1}  It turned out that the values of residual
potential difference depend on separation which means that, in addition to the
Casimir force and well understood electric forces due to the applied potentials,
there were some uncontrolled electrostatic forces due to surface patches.\cite{4,8,9}

Contrary to many experiments on measuring the Casimir force,\cite{1,6,8,9,13}
the Casimir metrology platform\cite{15} does not provide
the means for an independent
 measurement of the sphere-plate separations.
 The absolute separations are determined
 from the fit of the measurement data to two versions of the theory, i.e., to
 the zero-temperature Casimir force between the ideal metal sphere and plate
and to the perturbation expansion of the Casimir force in two small
parameters\cite{22}
(the relative temperature and relative penetration depth).
Surprisingly, it was found that
an ideal metal model at zero temperature leads to better agreement with the
measured data than the perturbation expansion taking into account corrections due to nonzero temperature and nonideality of metals (the root-mean-square
deviations equal to 7.4\,pN and 10.5\,pN, respectively).
This result is in contradiction with all previous precision measurements
of the Casimir force including the first experiment of this kind\cite{22a}
performed in 1998.

It should be taken into account, however, 
that the  perturbation expansion used
is applicable only at separations exceeding several hundred nanometers\cite{22}
and cannot be compared with the measurement data of Ref.~\citen{15} which are
taken at separations down to 65\,nm. 
At such short separations one should perform
numerical computations by substituting the optical data for the complex index
of refraction of boundary metals into the Lifshitz formula.

In Fig.~1 we present the computational results for the Casimir force between
the Ag sphere of $R=55\,\mu$m radius (as in Ref.~\citen{15}) and Au plate as
functions of separation obtained using the Lifshitz formula
at $T=300\,$K (the bottom solid
line), using the perturbation expansion\cite{22} (the dashed line) and
assuming the ideal-metal sphere and plate at zero
temperature (the top solid line).
Computations are performed taking into account the roughness of the plate
and sphere surfaces\cite{22R}
(with the root-mean-square amplitudes equal to 2 and
8\,nm, respectively\cite{15}).
We note that at separtions below 200\,nm the computational results obtained
using the extrapolations of the optical data to zero frequency by means of the
plasma and Drude models are rather close to each other and cannot be
discriminated in this experiment.
As is seen in Fig.~1, the ideal-metal Casimir
force deviates significantly from the accurate theory at all separations
below 200\,nm, whereas the perturbation expansion is in a rather good
agreement with it already at $a>130\,$nm. From Fig.~1 we conclude that the
largest measured force point\cite{15} ($F_C=635.5\,$pN) was obtained not at
the absolute separations of $a=65$ or 63\,nm (as claimed in Ref.~\citen{15}
from the fit to ideal metal  or perturbation theories, respectively),
but at some separation below 50\,nm.
\begin{figure}[t]
\vspace*{-9.cm}
\centerline{\hspace*{-0.1cm}\includegraphics[width=6.50in]{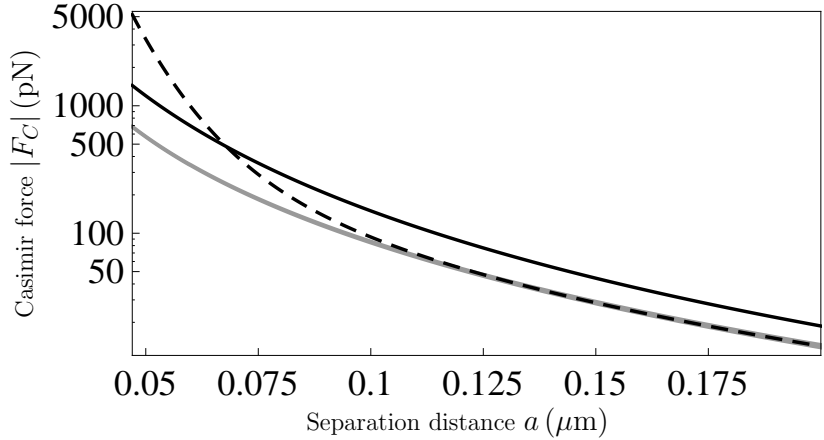}}
\vspace*{-7.9cm}
\caption{The Casimir force between an Ag sphere and an Au plate at $T=300\,$K
is shown as a function of separation when computations are performed using
the Lifshitz formula and the optical data of both metals (the bottom solid line),
using the perturbation theory in the relative penetration depth and relative
temperature (the dashed line) and using the idealization of ideal metal at
zero temperature (the top solid line).
\protect\label{fg1}}
\end{figure}

One should note also that a minimization of the root-mean-square deviation
between the data points and theoretical predictions is not an appropriate
method when measuring the strongly nonlinear quantities\cite{22b}
 such as the Casimir
force. As was shown long ago\cite{23} this method leads to quite different
results when it is used within different separation intervals.
It has been known that Sparnaay\cite{24} spelled out three fundamental
requirements necessary for performing precise and reproducible
measurements of the Casimir force.
According to one of these requirements, precise
independent and reproducible determination of the
separation between the test bodies must be performed in any
Casimir experiment.

\section{Conclusions}

As is seen from the foregoing, in the last four years great interest has
been expressed to measuring the Casimir force in different configurations and
to applications of this force in nanotechnology. Here we discussed only two
experiments which have problems in the comparison between experiment and theory.
Until the present time the Casimir interaction between superconductors has not
been measured. Because of this first measurement\cite{14} of the differences
in Casimir pressures when decreasing temperature
from above to below $T_c$ is undeniably
interesting. In future it is desirable to measure the absolute Casimir
pressures between superconductors both above and below the
critical temperature and compare the obtained results with different
theoretical predictions.

The use of a commercial capacitive sensor for demonstration of the Casimir
force in ambient conditions\cite{15} is also promising for various applications.
However, to take a status of the Casimir metrology platform, this work should be
supplemented  with an independent measurement of the sphere-plate
separations and  compared with the proper theory.

In the near future one could expect also the realization of proposed
experiments\cite{11,12,18} on measuring the Casimir forces between parallel
plates and superconductors, as well as the universal experiments aimed to
find out how free charge carriers influence the Casimir force between
metallic and semiconductor materials.\cite{25,26}

\section*{Acknowledgments}
V.~M.~M.~was partially funded by the Russian Foundation for Basic
Research, Grant No. 19-02-00453 A. His work was also partially supported by the
Russian Goverment Program of Competitive Growth of Kazan Federal University.

\end{document}